\documentclass[
preprintnumbers,
preprint,
a4paper,
showpacs,
amsmath,
amssymb,
byrevtex
]{revtex4}
\usepackage{amsmath,amssymb}
\usepackage{graphicx}
\begin{document}
\title{Black ring formation in particle systems}
\author{Hirotaka Yoshino}
\email{yoshino@gravity.phys.nagoya-u.ac.jp}
\author{Yasusada Nambu}
\email{nambu@gravity.phys.nagoya-u.ac.jp}
\affiliation{Department of Physics, Graduate School of Science, Nagoya
University, Chikusa, Nagoya 464-8602, Japan}
\preprint{DPNU-04-09}
\date{\today}
\begin{abstract}
It is known that the formation of  
apparent horizons with non-spherical topology is possible
in higher-dimensional spacetimes. One of these is the black ring
horizon with $S^1\times S^{D-3}$ topology where $D$ is the
spacetime dimension number.  In this paper,
we investigate the black ring horizon formation in systems
with $n$-particles. 
We analyze two kinds of system: the high-energy $n$-particle system 
and the momentarily-static $n$-black-hole initial data. 
In the high-energy particle system, we prove that the black ring
horizon does not exist at the instant of collision for any $n$. 
But there remains a possibility that the black ring forms after
the collision and this result is not sufficient. Because calculating
the metric of this system after the collision is difficult, we
consider the momentarily-static 
$n$-black-hole initial data that can
be regarded as a simplified $n$-particle model and numerically solve the
black ring horizon that surrounds all the particles.
Our results show that there is the minimum particle number that is
necessary for the black ring formation and this number depends on $D$.
Although many particle number is required in five-dimensions,
$n=4$ is sufficient for the black ring formation in the $D\ge 7$ cases. 
The black ring formation
becomes easier for larger $D$. 
We provide a plausible physical interpretation of our results
and discuss the validity of 
Ida and Nakao's conjecture for the horizon formation
in higher-dimensions.
Finally we briefly discuss 
the probable methods of producing the black rings in accelerators. 
\end{abstract}
\maketitle
\section{Introduction}

There have been two recent proposals known as the brane world scenarios
(the ADD scenario~\cite{ADD98} and the RS scenario~\cite{RS99}), 
which attempt to solve the hierarchy problem. 
In these scenarios, our three-dimensional space is the 3-brane embedded 
in extra-dimensions. 
In the ADD scenario, the extra-dimensions are flat and 
far larger compared to the Planck length. Accordingly
the Planck energy $M_P$ determined by the higher-dimensional gravitational 
constant $G_D$ becomes smaller than the four-dimensional Planck energy 
$M_4=10^{18}{\rm GeV}$, and it could be as low as $O(\rm{TeV})$. 
In the RS scenario, there are two branes with positive and negative tension
respectively in a five-dimensional spacetime, 
and the bulk is filled with a negative cosmological constant
$\Lambda$.
Due to this negative $\Lambda$, the extra-dimension has a warped configuration
whose effective volume for observers in the negative tension brane
becomes large and this also leads to the small Planck energy.

If the Planck energy is at $O({\rm TeV})$, 
the properties of the black holes smaller than
the scale of the extra-dimensions are totally altered. A black hole
is centered on the brane but is extending out into the large
extra-dimensions. Hence it becomes a higher-dimensional black hole.
Moreover, its gravitational radius is far larger than the four-dimensional
gravitational radius with the same mass. It was pointed out that
the black hole production in the future-planned accelerators would be
possible in this case~\cite{BF99, GT01, DL01}. 
In the particle collision with the energy above
the Planck scale, the gravity becomes the dominant interaction
and the semi-classical treatment is 
expected to become valid, although recently a counter-statement
which is based on the generalized uncertainty principle has been
proposed~\cite{CDM03}. 
If we assume that the maximal impact parameter for the black hole production
has the same order as
 the gravitational radius corresponding to the central energy,
the cross section becomes sufficiently large to produce one black hole
per one second at the Large Hadron Collider (LHC) in CERN. 
Because such black holes rapidly evaporate due to the Hawking radiation
and they radiate mainly on the brane~\cite{EHM00}, 
we are able to detect
the signals of the produced black holes at accelerators. 
The behavior of the produced black holes  
and the expected signals are first discussed by 
Dimopoulos and Landsberg~\cite{DL01}
and Thomas and Giddings~\cite{GT01},
and subsequently a great amount of papers concerning the 
black holes at accelerators appeared.

One of the necessary investigations to improve the
experimental estimates would be to analyze
the black hole formation in the high-energy particle collisions
in higher-dimensions. 
As a first step, the approximation neglecting
the tension of the brane and the structure of the extra-dimensions
would be appropriate if we consider
the case where the central energy of the particle system is
far above the Planck scale and the gravitational radius of the system
is smaller compared to the characteristic scale of the extra-dimensions. 
Eardley and Giddings analytically investigated the apparent 
horizon formation in this process in the four-dimensional 
spacetime~\cite{EG02}. Subsequently
we investigated the head-on collision in higher-dimensions
analytically~\cite{YN02} and  
the grazing collision in higher-dimensions with numerical 
calculations~\cite{YN03}.
Our results have shown that the cross section for the black hole production
is somewhat larger compared to the previous estimates. 

The above investigations were concerning the apparent horizons with
spherical topology. However it is known that the black hole horizon
topology in higher-dimensions are not restricted only to 
spherical topology.
Emparan and Reall generalized the Weyl solution and one of the solution
they derived was a
five-dimensional, asymptotically-flat, static black hole with horizon topology 
$S^1\times S^2$~\cite{ER02}, although 
this spacetime has a conical singularity that
supports the ring horizon and prevents it from collapsing. 
Subsequently they found the
rotating version of this black hole solution
without the conical singularity, which
was named the black ring~\cite{ER02_2}. 
Hence in the five-dimensional spacetime, the 
black holes are not only the generalized Kerr black holes found
by Myers and Perry~\cite{MP86}, and there is 
no uniqueness theorem for stationary black hole solutions
analogous to the four-dimensional case (see \cite{GIS02} for static cases).  
Although other solutions with non-spherical topology
horizons have not been found, many such solutions are expected to exist.
Ida and Nakao~\cite{IN02} discussed the possible horizon topology 
using the correlation between the power of the Newtonian potential
and the apparent horizon formation. According to their discussion,
the apparent horizon with torus topology $T^{D-2}=S^1\times\cdots\times S^1$ 
would be prohibited for all the spacetime dimension numbers $D$. 
However various topology horizons 
such as $S^1\times S^{D-3}$ would be allowed to form. 
Because these black holes with non-spherical horizon topology
have features that are characteristic to the higher-dimensional gravity,
whether we can produce such black holes would be an interesting theme. 
In this paper, we would like to analyze the formation of the black ring
horizon (i.e. the horizon with topology $S^1\times S^{D-3}$)
in particle systems.  

Whether the black ring can become the final
state depends on its stability. 
There are many evidences that the black rings are unstable.
First, if the angular momentum of the black ring is sufficiently large
(i.e. the parameter $\nu$ in \cite{ER02_2} is close to zero),
the Gregory-Laflamme instability~\cite{GL93} occurs
because the radius of the ring around $S^2$ becomes small
while the radius around $S^1$ becomes large~\cite{ER02_2}.
Next, due to the fact that the sequence of the black ring solution has the
minimum value of the angular momentum normalized by an 
appropriate power of the mass, 
the black ring would be unstable around this configuration  
because if one put some mass into the black ring, the mass 
would increase and the angular momentum would not change and thus there is no
black ring solution that corresponds to this situation~\cite{E02}. 
Finally, the rotating matter with ring configuration is not stable 
in the higher-dimensional Newtonian gravity~\cite{E02,IOP03}. 
If one write the effective 
potential as a function of the ring radius, there is only one unstable point
and is no stable point for $D\ge 6$, 
and there is only a marginally stable configuration
for $D=5$. All these situations support the instability
of the black rings for all the parameter range, although the explicit analysis
of the black ring perturbation has not been done. 
However, even if we admit 
the instability of the black ring, the black ring might form 
and exist until it changes to the other final state.
Ida, Oda, and Park~\cite{IOP03} discussed how the black ring evolves
if it forms in a high-energy two-particle system and if it traps almost all the
energy and the angular momentum of the system using the Newtonian 
approximation.  
According to their discussion, such a black ring increases its
radius around $S^1$ and expands until the Gregory-Laflamme
instability occurs. If it really happens, we might be able to 
observe some signals from the black rings in accelerators.

In this paper, we investigate the black ring formation
in $n$-particle systems for $n\ge 2$, because intuitively the black rings would
be easier to form in the system with more particles.
We would like to derive the condition for the black ring formation
in terms of the particle number $n$ and the location of particles. 
We analyze two kinds of system: the high-energy
particle system which we previously used in \cite{YN03} and
the $n$-body black hole initial data. 
First we will begin with the analysis of the high-energy particle systems.
The metric of the spacetime with one high-energy particle
has been written down by Aichelburg and Sexl~\cite{AS71} 
by boosting the Schwarzschild
black hole to the speed of light fixing the energy.
The resulting spacetime is a massless point particle accompanied by
a gravitational shock wave which distributes in the transverse
direction of motion. The high-energy two-particle system can be introduced
by combining two shock waves~\cite{DEP92}. In our previous paper~\cite{YN03},
we analyzed the apparent horizon on the union 
of two incoming shock waves (i.e. at the instant of collision)
and showed that the black hole horizon (i.e. the horizon with
topology $S^{D-2}$) forms. 
In this paper, we will introduce
the equation for the black ring horizon on this slice. Unfortunately,
we can easily show that there is no solution of the black ring horizon
on this spacetime slice.
Intuitively, it is natural to expect that the black ring would form
in a high-energy $n$-particle system if $n$ is sufficiently large. 
But also in the $n$-particle system, we can show that there is
no black ring horizon on the union of $n$ incoming shock waves. 
These results don't mean that the black ring cannot form
in high-energy particle systems: 
they just mean that there is no black ring horizon
at the instant of collision and there remains the possibility of its
formation after the collision of the shocks. To obtain the rigorous answer,
we should analyze the spacetime slice after the collision of the shocks.
However, this would be extremely difficult because the shocks
nonlinearly interact after the collision 
and no one has succeeded to calculate the
metric after the collision with the impact parameter.
Instead, we next analyze the more simplified situation, 
the conformally-flat
momentarily-static (i.e. time-symmetric) $n$-black-hole initial data 
originally introduced by Brill and Lindquist 
in the four-dimensional case~\cite{BL63}.
This initial data also can be regarded as a $n$-particle system
because it is the solution of the Hamiltonian constraint of the
Einstein equation with $\delta$-function sources.
Although this setting might be too simplified especially because
each particle is not moving, this initial data provides the $n$
particles that are interacting each other and we would be able to
obtain a lot of implications about 
the black ring formation in higher-dimensions. We
numerically solve the apparent horizon with $S^1\times S^{D-3}$ topology
that surrounds all the particles, 
and derive the minimum number $n_{\rm min}$ 
that is necessary for the black ring formation 
for each spacetime dimension number $D$. 
Using these results, we would like to discuss the feature of the 
black ring formation in higher-dimensions. 

This paper is organized as follows. In Sec. II, we investigate 
the high-energy particle system and show that the black ring horizon
does not exist at the instant of collision. In Sec. III, we
analyze the momentarily-static $n$-black-hole system.
We explain the numerical methods of finding apparent horizons
with $S^1\times S^{D-3}$ and $S^{D-2}$ topology, 
and show the numerical results.
We also provide a plausible physical interpretation for our results
and discuss the validity of Ida and Nakao's conjecture~\cite{IN02} 
for the apparent horizon formation in higher-dimensions. 
Sec. IV is devoted to summary and discussion. The brief discussion 
concerning the probable methods of black ring production in accelerators
is also included.

\section{High-energy particle system}

In this section, we investigate the existence of the black ring horizon
in the high-energy $n$-particle system. 
First we analyze the two-particle system,
in which we previously investigated the black hole formation~\cite{YN03}. 
The set up of the system
is as follows. To obtain the metric of a high-energy one-particle system, 
we boost the $D$-dimensional Schwarzschild
solution to the speed of light. 
The Schwarzschild metric is given by 
\begin{equation}
ds^2=-\left(1-\frac{16\pi G_DM}{(D-2)\Omega_{D-2}r^{D-3}}\right)dt^2+\left(1-\frac{16\pi
    G_DM}{(D-2)\Omega_{D-2}r^{D-3}}\right)^{-1}dr^2
+r^2d\Omega_{D-2}^2, 
\end{equation}
where $d\Omega_{D-2}^2$ and $\Omega_{D-2}$ are the line element and
the $(D-2)$-area of the $(D-2)$-dimensional unit sphere respectively,
and the horizon is located at $r=r_h(M)$ where  
\begin{equation}
r_h(M)\equiv 
\left(\frac{16\pi G_DM}{(D-2)\Omega_{D-2}}\right)^{1/(D-3)}.
\label{eq:Schwarzschild_radius}
\end{equation}
By boosting this metric in $z$-direction and taking a lightlike limit
fixing the energy $\mu=M\gamma$,
we obtain the Aichelburg-Sexl metric~\cite{AS71}
\begin{equation}
ds^2=-d\bar{u} d\bar{v}+\sum_{i=1}^{D-2}d\bar{x}_i^2+\Phi(\bar{x}_i)
\delta(\bar{u})d\bar{u}^2,
\label{eq:delta}
\end{equation}
where $\bar{u}=\bar{t}-\bar{z}$ and $\bar{v}=\bar{t}+\bar{z}$.
$\Phi$ depends only on the transverse radius 
$\bar{\rho}=\sqrt{\sum_{i=1}^{D-2}\bar{x}_i^2}$ and takes the form
\begin{equation}
\Phi(\bar{x}_i)=\left\{
\begin{array}{cc}
-8G_4\mu\log\bar{\rho},&\quad\text{for}~D=4, \\
{16\pi\mu G_D}/{\Omega_{D-3}(D-4)\bar{\rho}^{D-4}},&\quad\text{for}~D>4.\\
\end{array}
\right.
\end{equation}
The metric \eqref{eq:delta} is flat except $\bar{u}=0$ and 
the $\delta$ function in eq.~\eqref{eq:delta} indicates that 
there is a gravitational shock wave propagating in the
flat spacetime at the speed of light.
Mathematically, the origin of this $\delta$ function is due to the
fact that the 
two flat-spacetime coordinate systems $(\bar{u}, \bar{v}, \bar{x}_i)$ 
are discontinuously connected at $\bar{u}=0$. The continuous
coordinate system $(u, v, x_i)$ can be introduced by 
\begin{align}
& \bar{u}=u, \\
& \bar{v}=v+\Phi({x_i})\theta(u)
+\frac{u}{4}\theta(u)
\sum_k\left(\frac{\partial \Phi(x_i)}{\partial x_k}\right)^2,\\
& \bar{x}_j=x_j+\frac{u}{2}\theta(u)
\left(\frac{\partial\Phi({x_i})}{\partial x_j}\right), 
\end{align}
where $\theta(u)$ is the Heaviside step function. 
In this coordinate system, the metric is represented without the $\delta$
function:
\begin{equation}
ds^2=-dudv+H_{ik}H_{jk}dx_idx_j,
\end{equation}
where
\begin{equation}
H_{ij}=\delta_{ij}+\frac{u}{2}\theta(u)
\frac{\partial^2\Phi}{\partial x_i\partial x_j}.
\end{equation}
We can superpose two solutions to obtain the exact geometry outside
the future light cone of the collision of the shocks:
\begin{align}
&ds^2=-dudv+\left(H^{(+)}_{ik}H^{(+)}_{jk}+H^{(-)}_{ik}H^{(-)}_{jk}
-\delta_{ij}\right)dx^idx^j,
\end{align}
where
\begin{align}
&H^{(+)}_{ij}=\delta_{ij}+\frac{u}{2}\theta(u)
\frac{\partial^2}{\partial x_i\partial x_j}
\Phi^{(+)}(\boldsymbol{x}),\\
&H^{(-)}_{ij}=\delta_{ij}+\frac{v}{2}\theta(v)
\frac{\partial^2}{\partial x_i\partial x_j}
\Phi^{(-)}(\boldsymbol{x}).
\end{align}
Here $\boldsymbol{x}$ is the point in the $(D-2)$-space $(x_i)$
and $\Phi^{(\pm)}(\boldsymbol{x})$ are defined as 
$\Phi^{(\pm)}(\boldsymbol{x})\equiv\Phi(\boldsymbol{x}-\boldsymbol{x}_{\pm})$
where $\boldsymbol{x}_{\pm}$ denote the locations of the incoming
particles in this subspace and we set 
\begin{equation}
\boldsymbol{x}_{\pm}=(\pm b/2,0,...,0),
\end{equation}
where $b$ is the impact parameter.
This system was originally introduced by D'Eath and Payne in the 
four-dimensional case~\cite{DEP92}.

Eardley and Giddings~\cite{EG02} derived the equation of the apparent horizon
on the union of the incoming shock waves, i.e., $v\le 0, u=0$ and 
$u\le 0, v=0$.
We briefly review this method because it is useful also for the
black ring horizon case. 
They assumed that the horizon is given by the union of $(D-2)$-surfaces
$\mathcal{S}_+$ in $v\le 0, u=0$ and $\mathcal{S}_-$ in $u\le 0, v=0$
that are given by $v=-\Psi^{(+)}(x_i)$ and $u=-\Psi^{(-)}(x_i)$ respectively.
The surfaces $\mathcal{S}_{\pm}$ are combined at $(D-3)$-surface 
$\mathcal{C}$ in $u=v=0$.  
Because there is a symmetry in this system, we have only to consider
the $\mathcal{S}_+$ surface. 
Although the function $\Psi^{(+)}$ diverges at 
$\boldsymbol{x}=\boldsymbol{x}_{+}$, the surface is continuous at that point
if one see in the $(\bar{u}, \bar{v}, \bar{x}_i)$ coordinate.
In this coordinate, the surface $\mathcal{S}_+$ is located at
\begin{equation}
\bar{v}=\Phi^{(+)}(x_i)-\Psi^{(+)}(x_i)\equiv h(\bar{x}_i),
\end{equation}
where $x_i=\bar{x}_i$ is satisfied on $u=0$.
The outgoing null normal for this surface is given by
\begin{equation}
k^a=\left(1, 
\frac14\sum_k\left(\frac{\partial h}{\partial \bar{x}_k}\right)^2,
\frac12\left(\frac{\partial h}{\partial \bar{x}_i}\right)\right),
\label{eq:normal}
\end{equation}
and the condition for zero expansion becomes
\begin{equation}
\bar{\nabla}^2_f h=0,
\label{eq:Laplace}
\end{equation}
where $\bar{\nabla}_f^2$ denotes the flat space Laplacian
in the $(\bar{x}_i)$ coordinate. 
To find the black hole horizon, one should calculate the
continuous and smooth solution of eq.~\eqref{eq:Laplace} with two
boundary conditions imposed at $\mathcal{C}$ (see \cite{EG02} for
details).   

\begin{figure}[tb]
\centering
{
\includegraphics[width=0.5\linewidth]{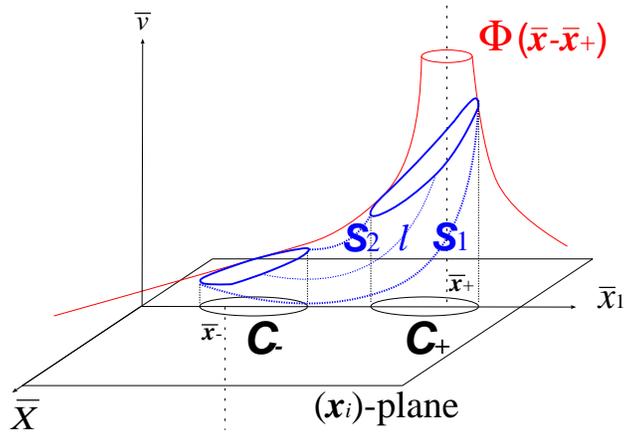}
}
\caption{\label{Laplace2}
The schematic shape of the solution for the ring horizon
that we would like to construct. The ring horizon becomes
a double-valued function of $(\bar{x}_1, \bar{X})$ where
$\bar{X}$ denotes the distance from the $\bar{x}_1$-axis
in the $(D-2)$-space $(\bar{x}_i)$. 
There are two boundaries denoted by $\mathcal{C}_{\pm}$, which
surround $\boldsymbol{x}_{\pm}$ respectively. We divide the surface into
$\mathcal{S}_1$ (the lower one) and $\mathcal{S}_2$ (the upper one),
and $l$ is the location where the two surfaces are connected
and the gradient diverges. Unfortunately we can show that the
solution like this figure does not exist.
}
\end{figure}

Now we consider whether the black ring horizon with $S^1\times S^{D-4}$
exists on the same slice. Due to this horizon topology, 
the location of the horizon $\mathcal{S}_+$ becomes a double-valued
function of $(\bar{x}_i)$. We divide $\mathcal{S}_+$ into two parts
denoted by 
\begin{equation}
\mathcal{S}_1:~\bar{v}=h_1(\bar{x}_i),~~\textrm{and}~~\mathcal{S}_2:~\bar{v}=h_2(\bar{x}_i),
\end{equation}
where we impose $h_2\ge h_1$. 
Using the symmetries of this system, $h_1$ and $h_2$
depends only on $\bar{x}_1$ and 
$\bar{X}\equiv\sqrt{\sum_{k=2}^{D-2}\bar{x}_k^2}$. 
The shape of the ring horizon
that we expect to exist is shown in Fig.~\ref{Laplace2}.
$\mathcal{S}_1$ and $\mathcal{S}_2$ are combined at a $(D-3)$-surface $l$,
where the gradient of $h_1$ and $h_2$ diverges. There are two
boundaries $\mathcal{C}_+$ and $\mathcal{C}_-$, where we should
impose boundary conditions. The null normal of the surface $\mathcal{S}_2$
becomes
\begin{equation}
k^a=\left(0,1,0,...,0\right),
\end{equation}
and clearly $\mathcal{S}_2$ has zero expansion for all $h_2$.
The null normal of $\mathcal{S}_1$ and the equation for 
zero expansion coincide with
eq.~\eqref{eq:normal} and eq.~\eqref{eq:Laplace} respectively
if we put $h\equiv h_1$. 
However, we can easily show the non-existence of
the solution $h$ that satisfies the above conditions.
Assuming the existence of the solution, 
the location of $l$ is expressed as 
\begin{equation}
\bar{X}=f(\bar{x}_1),~~\bar{v}=g(\bar{x}_1).
\end{equation}
Because the surface $\mathcal{S}_+$ should be smooth, 
the surface $\mathcal{S}_1$ is well-approximated by
\begin{equation}
\bar{X}\simeq f(\bar{x}_1)-a(\bar{x}_1)
\left(h(\bar{x}_1,\bar{X})-g(\bar{x}_1)\right)^2,
\end{equation} 
in the neighborhood of 
the $(D-3)$-surface $l$ with some function $a(\bar{x}_1)$. 
Calculating $\bar{\nabla}^2 h$, we obtain
\begin{equation}
\bar{\nabla}^2 h\simeq 
-\frac{1+(\partial f/\partial \bar{x}_1)^2}{4a^2}
\left(h-g(\bar{x}_1)\right)^{-3}.
\label{eq:diverge}
\end{equation}
If we take a limit $h\to g(\bar{x}_1)$, the value of $\bar{\nabla}^2h$
diverges for any choice of $a(\bar{x}_1)$, $f(\bar{x}_1)$, and $g(\bar{x}_1)$,
and this shows that the surface like Fig.~\ref{Laplace2}
cannot satisfy the Laplace equation.
This result is due to the fact that if there is 
a $(D-3)$-dimensional singularity of the $(D-2)$-dimensional 
Laplace equation, both values of the solution and its gradient
should be finite in the neighborhood of the singularity. 
Although we have not considered the boundary conditions in this discussion,
this is sufficient for the proof of the non-existence of the solution
because satisfying eq.~\eqref{eq:Laplace} is a necessary condition
and  eq.~\eqref{eq:diverge} shows that the necessary condition
cannot be satisfied. 
Hence the black ring horizon does not exist on the union of
two incoming shock waves, although the black hole horizon 
can be calculated~\cite{YN03}.

Intuitively, it would be obvious that the black ring forms in a $n$-particle
system for large $n$. 
We can write down the metric of the high-energy $n$-particle
system outside the future light cone of the shock collisions similarly
to the two-particle case. However, one can easily see that
the black ring horizon does not exist on the union of $n$ incoming shocks,
simply because the horizon equation is the $(D-2)$-Laplace equation
also in this system and it does not allow the divergence of
the gradient at the $(D-3)$-singularity. 
Hence we cannot find the ring horizon even in the $n$-particle
system. However, these results just show that the ring horizon does not exist
at the instant of collision and it might form after the collision. 
To obtain the answer, we should investigate the spacetime slice
after the collision of the shocks. This analysis is rather difficult
because the shocks nonlinearly interact each other after the collision
and no one has calculated the metric after the collision. 
Hence at this point, we don't understand anything about the
black ring formation in particle systems. 
To obtain some implications,
we consider a rather simplified situation in the next section:
the momentarily-static $n$-black-hole initial data.

\section{$n$ black hole system}

As we mentioned in previous sections, we consider the momentarily-static
$n$-black-hole initial data as a simplified model that we can analyze. 
This initial data provides $n$-particles that 
are interacting each other,
although each particle is not moving. By analyzing the formation of
the black ring horizon in this system, we are able to understand the feature 
of the black ring formation in $n$-particle systems in higher-dimensions.

\subsection{The Brill-Lindquist initial data}

Let $(\mathcal{M}, g_{ij},K_{ij})$ be the $(D-1)$-dimensional space
with the metric $g_{ij}$ and the extrinsic curvature $K_{ij}$. 
We assume the conformally-flat initial data of which metric is
\begin{equation}
ds^2=\Psi^{4/(D-3)}(\boldsymbol{x})
\left(dx^2+dy^2+dz^2+z^2d\Omega_{D-4}^2\right),
\end{equation}
where $\boldsymbol{x}$ denotes the location in the flat space and
each $(D-4)$-sphere is spanned by a $(\chi_1,...,\chi_{D-4})$ coordinate. 
We consider the momentarily static case, where
the extrinsic curvature becomes $K_{ij}=0$.
The Hamiltonian constraint becomes
\begin{equation}
\nabla_f^2\Psi=0,
\label{Hamiltonian}
\end{equation}
where $\nabla_f^2$ denotes the Laplacian of the $(D-1)$-dimensional
flat space. We select the solution of eq.~\eqref{Hamiltonian} 
with $n$ point sources:
\begin{equation}
\Psi=1+\sum_{k=1}^{n}\frac{4\pi G_D M/n}{(D-2)\Omega_{D-2}r_k^{D-3}},
\label{conformal_factor}
\end{equation}
where $M$ is the ADM mass and $r_k\equiv |\boldsymbol{x}-\boldsymbol{x}_k|$ 
is the distance between the point $\boldsymbol{x}$ and
the location of the point source $\boldsymbol{x}_k$ in the flat space. 
In the case of $n=1$ and $D=4$, this solution provides the Einstein-Rosen
bridge. The solution of $n\ge 2$ is the $n$ Einstein-Rosen bridges
originally introduced by Brill and Lindquist~\cite{BL63}.
Each $\boldsymbol{x}_k$ corresponds to the asymptotically
flat region beyond the Einstein-Rosen bridge. 
We locate the $n$ point sources at $n$ vertexes of a regular polygon
that is inscribed in a circle with the radius $R$.
More precisely, we set each $\boldsymbol{x}_k$ as
\begin{equation}
\boldsymbol{x}_k=
\left(R\cos\frac{2\pi}{n}(k-1), R\sin\frac{2\pi}{n}(k-1), 0\right),
\label{location}
\end{equation}
in the $(x,y,z)$ coordinate. 
As we see from the fact that eq.~\eqref{conformal_factor}
is a solution of the Hamiltonian constraint~\eqref{Hamiltonian} with $n$
point sources, this initial data can be regarded as a $n$-point-particle
system. Note that the parameters that specify the initial data 
are the spacetime dimension number $D$, the particle number $n$,
and the parameter $R$ to locate the particles in eq.~\eqref{location}.

\subsection{Methods of finding apparent horizons}

Now we explain the methods of finding apparent horizons.
Because this space is time-symmetric, the equation for the
apparent horizon is
\begin{equation}
\theta_+=\nabla_is^i=0,
\end{equation}
where $\theta_+$ stands for the expansion of outgoing null geodesic congruence,
$s^i$ is the unit normal of the apparent horizon and
$\nabla_i$ denotes the covariant derivative with respect to $g_{ij}$. 
We would like to find the black ring horizon with $S^1\times S^{D-3}$
topology. Because the apparent horizon is defined as the
outermost marginally trapped surface, we also should solve 
the black hole horizon with $S^{D-2}$ topology that surrounds 
all the particles to judge whether the black ring horizon 
is inside the black hole horizon or not.

We introduce a new coordinate $(\rho, \xi, \phi, \chi_1,...,\chi_{D-4})$
by
\begin{align}
&x=(R_0+\rho\cos\xi)\cos\phi,\\
&y=(R_0+\rho\cos\xi)\sin\phi,\\
&z=\rho\sin\xi.
\end{align}  
The metric becomes
\begin{align}
ds^2=\Psi^{4/(D-3)}
\left(d\rho^2+\rho^2d\xi^2+(R_0+\rho\cos\xi)^2d\phi^2+
\rho^2\sin^2\xi d\Omega_{D-4}^2\right).
\end{align}
This coordinate system is useful for solving the black ring horizon
with $S^1\times S^{D-3}$ topology, because we can specify
the location of the black ring horizon by $\rho=h(\xi,\phi)$
if we choose the value of $R_0$ appropriately. For this purpose, 
the line $\rho=0$ should be inside the black ring horizon.
The unit normal $s^i$ for this surface is given by
\begin{equation}
s^i=\frac{\Psi^{-2/{(D-3)}}}
{\sqrt{1+h_{,\xi}^2/\rho^2+h_{,\phi}^2/(R_0+\rho\cos\xi)^2}}
\left(1,\frac{-h_{,\xi}}{\rho^2},\frac{-h_{,\phi}}{(R_0+\rho\cos\xi)^2}
,0,...,0\right).
\end{equation}
Calculating $\theta_+=\nabla_is^i=0$, we derive the equation for
the black ring horizon as follows:
\begin{multline}
\left(1+\frac{h_{,\phi}^2}{(R_0+h\cos\xi)^2}\right)h_{,\xi\xi}
+\frac{\left(h^2+{h_{,\xi}^2}\right)}{(R_0+h\cos\xi)^2}
h_{,\phi\phi}
-\frac{2h_{,\xi}h_{,\phi}h_{,\xi\phi}}{(R_0+h\cos\xi)^2}
-\frac{h_{,\xi}^2}{h}\\
-\frac{(h\cos\phi+h_{,\xi}\sin\xi)}{(R_0+h\cos\xi)^3}hh_{,\phi}^2
-\left(h^2+{h_{,\xi}^2}+\frac{h^2h_{,\phi}^2}{(R_0+h\cos\xi)^2}\right)
\Bigg[
\frac{D-3}{h}
+\frac{(h\cos\xi+h_{,\xi}\sin\xi)}{h(R_0+h\cos\xi)}\\
-\frac{(D-4)\cot\xi}{h^2}h_{,\xi}
+\frac{2(D-2)}{(D-3)\Psi}
\left({\Psi_{,\rho}}
-\frac{h_{,\xi}\Psi_{,\xi}}{h^2}
-\frac{h_{,\phi}\Psi_{,\phi}}{(R_0+h\cos\xi)^2}\right)
\Bigg]
=0.
\label{eq:ring}
\end{multline}
Because this space has a mirror symmetry about the equatorial plane
and a discrete symmetry for the rotation about the $z$-axis, 
it is sufficient to solve in the range $0\le \xi\le \pi$
and $0\le \phi \le \pi/n$. 
We solved eq.~\eqref{eq:ring} under the boundary conditions
\begin{align}
&h_{,\xi}=0~~\textrm{at}~~\xi=0,\pi,\\
&h_{,\phi}=0~~\textrm{at}~~\phi=0,\pi/n,
\end{align}
using the SOR method. We selected the value of $R_0$ to be
$R_0=R$ for $n\ge 6$, $R_0=0.9R$ for $n=5$, $R_0=0.8R$ for $n=4$,
and $R_0=0.7R$ for $n=3$. 
The black ring horizon has the $(D-2)$-dimensional area 
$A_{D-2}^{\textrm{(BR)}}$, which is given by
\begin{equation}
A_{D-2}^{\textrm{(BR)}}=
2n\Omega_{D-4}\int_0^{\pi/n}d\phi\int_0^{\pi}d\xi
\Psi^{{2(D-2)}/{(D-3)}}(h\sin\xi)^{D-4}
\sqrt{(R_0+h\cos\xi)^2(h^2+h_{,\xi}^2)+h^2h_{\phi}^2}.
\end{equation}
The black ring horizon is the minimal surface
of this area $A_{D-2}^{\textrm{(BR)}}$.

Next we explain the method of finding the black hole horizon
with $S^{D-2}$ topology. For this purpose, it is useful to
choose the coordinate $(r,\theta,\phi,\chi_1,...,\chi_{D-4})$ where
\begin{align}
&x=r\sin\theta\sin\phi,\\
&y=r\sin\theta\cos\phi,\\
&z=r\cos\theta.
\end{align}
We give the location of the black hole horizon by $r=\tilde{h}(\theta,\phi)$.
The unit normal of this surface becomes
\begin{equation}
s^i=\frac{\Psi^{-2/{(D-3)}}}
{\sqrt{1+\tilde{h}_{,\theta}^2/r^2+\tilde{h}_{,\phi}^2/r^2\sin^2\theta}}
\left(1,\frac{-\tilde{h}_{,\theta}}{r^2},\frac{-\tilde{h}_{,\phi}}{r^2\sin^2\theta}
,0,...,0\right).
\end{equation}
The equation $\theta_+=0$ for the black hole horizon becomes
\begin{multline}
\left(1+\frac{\tilde{h}_{,\phi}^2}{\tilde{h}^2\sin^2\theta^2}\right)\tilde{h}_{,\theta\theta}
+\left(1+\frac{\tilde{h}_{,\theta}^2}{\tilde{h}^2}\right)\frac{\tilde{h}_{,\phi\phi}}{\sin^2\theta}
-\frac{2\tilde{h}_{,\theta}\tilde{h}_{,\phi}\tilde{h}_{,\theta\phi}}{\tilde{h}^2\sin^2\theta}\\
-\left(\tilde{h}_{,\theta}^2+\frac{\tilde{h}_{,\phi}^2}{\sin^2\theta}\right)\frac{1}{\tilde{h}}
+\frac{\tilde{h}_{,\theta}\tilde{h}_{,\phi}^2\cos\theta}{\tilde{h}^2\sin^3\theta}
-\left(1+\frac{\tilde{h}_{,\theta}^2}{\tilde{h}^2}+\frac{\tilde{h}_{,\phi}^2}{\tilde{h}^2\sin^2\theta}\right)
\Bigg[
(D-2)\tilde{h}\\
+\frac{2(D-2)}{(D-3)\Psi}
\left({\Psi_{,r}}\tilde{h}^2
-{\Psi_{,\theta}\tilde{h}_{,\theta}}
-\frac{\Psi_{,\phi}\tilde{h}_{,\phi}}{\sin^2\theta}\right)
-\left(\cot\theta-(D-4)\tan\theta\right)\tilde{h}_{,\theta}
\Bigg]=0.
\label{eq:spheri}
\end{multline}
We solved this equation under the boundary conditions
\begin{align}
&\tilde{h}_{,\theta}=0~~\textrm{at}~~\theta=0,\pi/2,\\
&\tilde{h}_{,\phi}=0~~\textrm{at}~~\phi=0,\pi/n.
\end{align}
Because there is a coordinate singularity at $\theta=0$,
a careful treatment is required for the numerical calculation.
We imposed the boundary condition at $\theta=0$ as follows.
First, expressing the location of the apparent horizon as $z=f(x,y)$,
we calculated the equation for the apparent horizon in terms of $f(x,y)$.
Next we wrote down the differential equation of $f(x,y)$ at $(x,y)=(0,0)$.
Finally we rewrote this differential equation in terms of $\tilde{h}$.
This differential equation at $\theta=0$ is somewhat complicated
especially in the cases where $n$ is a odd number. 
The $(D-2)$-dimensional
area of the black hole horizon is given as follows:
\begin{equation}
A_{D-2}^{\textrm{(BH)}}=
2n\Omega_{D-4}\int_0^{\pi/n}d\phi\int_0^{\pi/2}d\theta
\Psi^{{2(D-2)}/{(D-3)}}\tilde{h}^{D-3}\cos^{D-4}\theta
\sqrt{(\tilde{h}^2+\tilde{h}_{,\theta}^2)\sin^2\theta+\tilde{h}_{\phi}^2}.
\end{equation}
The black hole horizon is the minimal surface of 
this area $A_{D-2}^{\textrm{(BH)}}$.

We solved with $40\times 40$ grids for both horizons.
The error is estimated as follows. In $n=2$ cases, 
the spherical horizon can be solved using the Runge-Kutta method
with a fairy good accuracy because the space is axi-symmetric.
Comparing the two results, we evaluate the error to be about 1\%
in $n=2$ cases.
The error is expected to be smaller than this value for $n\ge 3$
because the parameter range of $\theta$ becomes smaller while
we use the same grid number.
To evaluate the error of the ring horizon, 
we solved with $80\times 80$ grids in some cases. The error is also 
about 1\% or smaller than this value for the ring horizon. 
Hence the possible changes of the values in Tables \ref{data:D5},
\ref{data:D6}, \ref{data:D7_8} and \ref{data:D9_10}
which we will show later
are expected to be within $0.01$.

\subsection{Numerical results}

Now we show the numerical results.
In this subsection, we use 
\begin{equation}
\tilde{r}_h\equiv \left(
\frac{4\pi G_D M}{(D-2)\Omega_{D-2}}
\right)^{1/(D-3)}
\end{equation}
as the unit of the length. In this unit,
the location of the apparent horizon in the $R=0$ cases becomes
$r=1$. 
We calculated the horizons for each $R$ with $0.01$ intervals.

\begin{table}[t]
\caption{ \label{data:D5}The values of $R_{min}^{\textrm{(BR)}}$, 
$R_{max}^{\textrm{(BR)}}$, and $R_{max}^{\textrm{(BH)}}$ 
for $n=4,..., 11$ in the $D=5$ case. 
The ring horizon does not form for $n\le 6$.
Because $R_{max}^{\textrm{(BR)}}<R_{max}^{\textrm{(BH)}}$ for
$n=7$ and $8$, the ring horizon is inside the black hole apparent horizon.
For the ring apparent horizon formation, $n\ge n_{\rm min}=9$ is necessary.}
\begin{ruledtabular}
\begin{tabular}{c|cccccccc}
  $n$ & 4 & 5 & 6 & 7 & 8 & 9 & 10 & 11 \\
  \hline 
  $R_{min}^{\textrm{(BR)}}$ & -- & -- & -- & 0.79 & 0.79 & 0.79
  & 0.79 & 0.79 \\
  $R_{max}^{\textrm{(BR)}}$ & -- & -- & -- & 0.82 & 0.86 & 0.91 
  & 0.95 & 0.99\\
  $R_{max}^{\textrm{(BH)}}$ & 0.84 & 0.87 & 0.88 & 0.89 & 0.90 & 0.90 
  & 0.90 &0.90\\
\end{tabular}
\end{ruledtabular}
\end{table}

\begin{figure}[t]
\centering
{
\includegraphics[height=0.25\textheight]{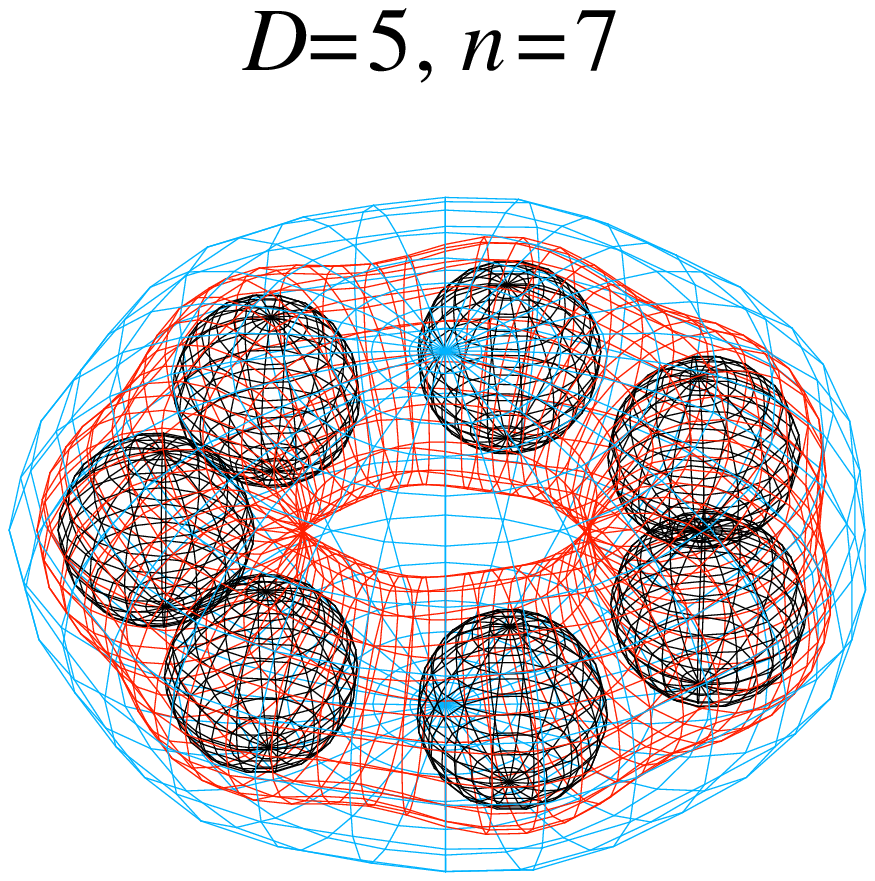}\hspace{10mm}
\includegraphics[height=0.25\textheight]{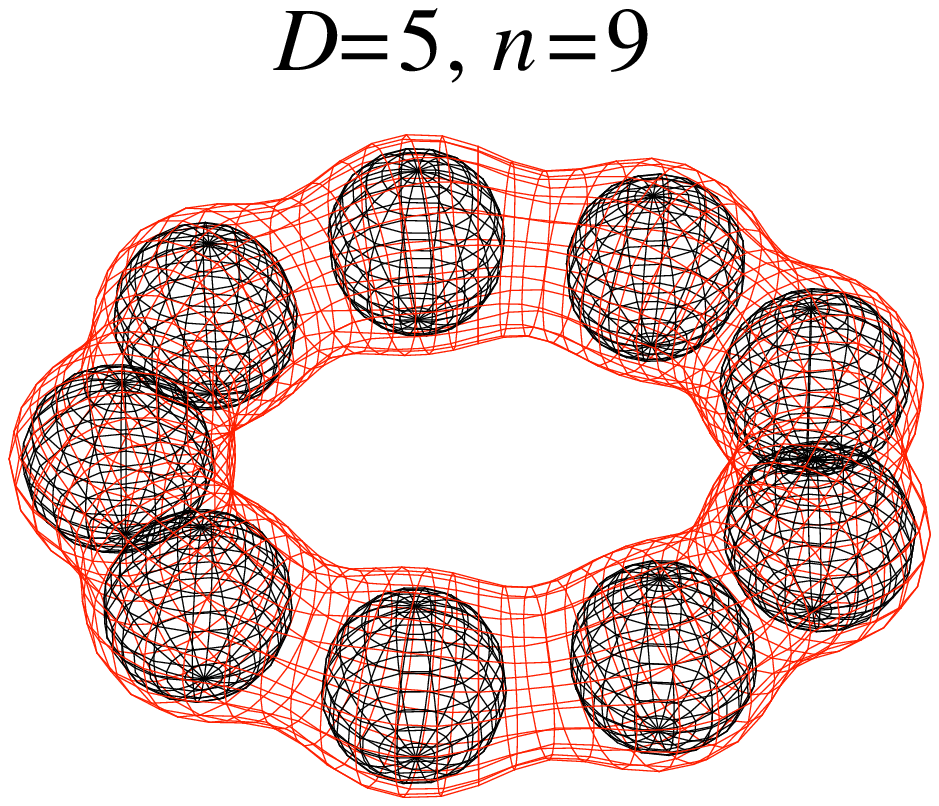}
}
\caption{\label{pic:D5}
The black ring horizon for $n=7$ ($R=0.8\tilde{r}_h$) and $n=9$ ($R=0.91\tilde{r}_h$)  
in five-dimensions.
Although the black ring horizon forms
in the $n=7$ case, it is inside the
black hole horizon and hence the system is actually a black hole.
For $n\ge 9$, the black ring horizon that is not enclosed by
the black hole horizon appears.}
\end{figure}

\begin{table}[t]
\caption{ \label{data:D6}The values of $R_{min}^{\textrm{(BR)}}$, 
$R_{max}^{\textrm{(BR)}}$, and $R_{max}^{\textrm{(BH)}}$ 
for $n=3,..., 7$ in the $D=6$ case. 
The ring horizon does not form for $n=3$
and the ring horizon is inside the black hole apparent horizon
for $n=4$.
For the ring apparent horizon formation, $n\ge n_{\rm min}=5$ is necessary.}
\begin{ruledtabular}
\begin{tabular}{c|ccccc}
  $n$ & 3 & 4 & 5 & 6 & 7  \\
  \hline 
  $R_{min}^{\textrm{(BR)}}$ & -- & 0.86 & 0.87 & 0.88 & 0.88 \\
  $R_{max}^{\textrm{(BR)}}$ & -- & 0.86 & 0.96 & 1.07 & 1.17 \\
  $R_{max}^{\textrm{(BH)}}$ & 0.83 & 0.86 & 0.88 & 0.89 & 0.89 \\
\end{tabular}
\end{ruledtabular}
\end{table}

\begin{figure}[t]
\centering
{
\includegraphics[height=0.25\textheight]{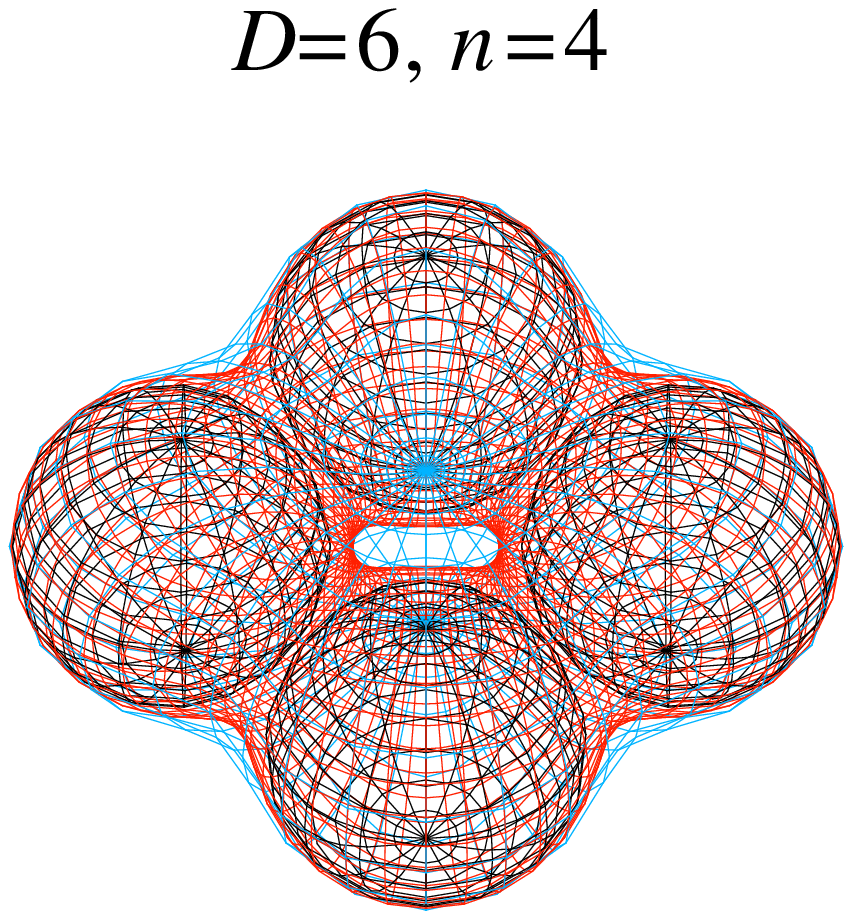}\hspace{10mm}
\includegraphics[height=0.25\textheight]{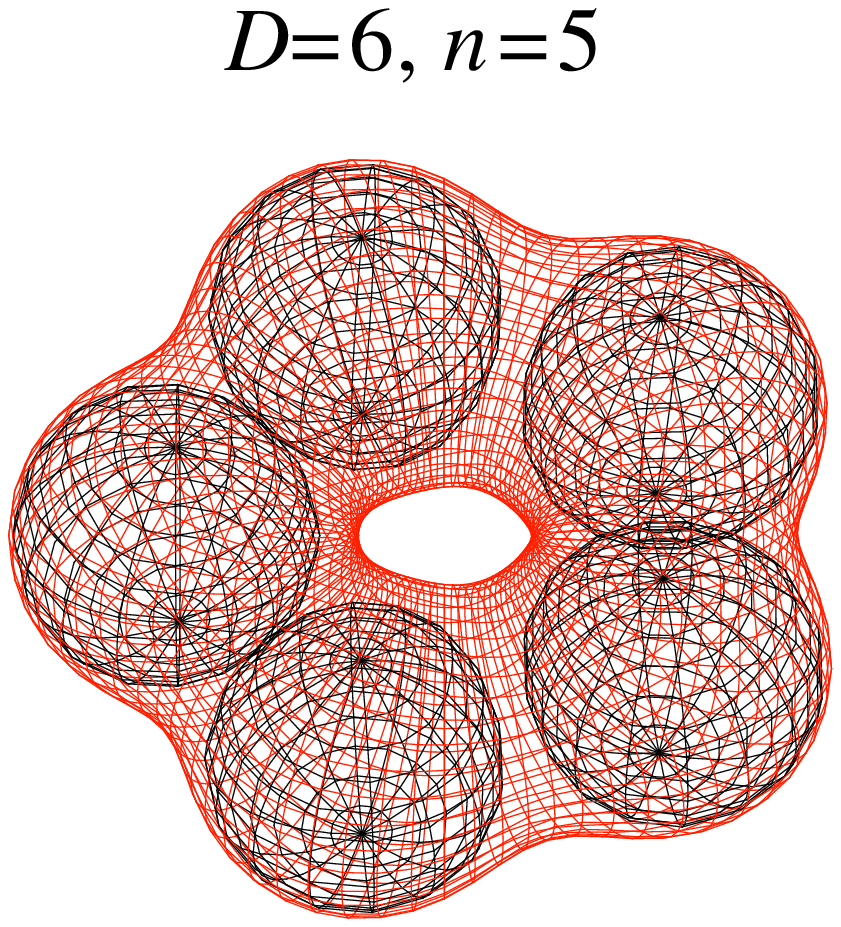}
}
\caption{\label{pic:D6}
The horizons in a four-particle system and a five-particle system in
six-dimensions ($R=0.86\tilde{r}_h$ and $R=0.9\tilde{r}_h$ respectively). 
In a four-particle system, the ring horizon is enclosed by a black hole horizon.
In a five-particle system, the ring horizon becomes the apparent horizon.}
\end{figure}

We begin with the $D=5$ case. In this case, 
we could find the black ring horizon for $n\ge 7$.
For each $n\ge 7$, there are two characteristic 
values $R_{min}^{\textrm{(BR)}}$
and $R_{max}^{\textrm{(BR)}}$ such that the black ring horizon
does not exist for $R<R_{min}^{\textrm{(BR)}}$ and $R>R_{max}^{\textrm{(BR)}}$,
and it exists for 
$R_{min}^{\textrm{(BR)}}\le R \le R_{max}^{\textrm{(BR)}}$. 
In the $R>R_{max}^{\textrm{(BR)}}$ cases, the throat of the
surface between neighboring two particles
becomes arbitrarily narrow (i.e. $h(\xi,\pi/n)$ becomes zero) 
during the iteration. This 
result shows that the distance between two particles 
neighboring each other should be sufficiently small for
the ring horizon formation. 
In the $R<R_{min}^{\textrm{(BR)}}$ cases, the hole of 
the surface becomes arbitrarily small during the iteration
(i.e. $h(\pi,\phi)$ approaches to $R_0$).
Hence the space between all the particles at the center
should be sufficiently large. 
Table~\ref{data:D5} shows the values of $R_{min}^{\textrm{(BR)}}$, 
$R_{max}^{\textrm{(BR)}}$, and $R_{max}^{\textrm{(BH)}}$, where $R_{max}^{\textrm{(BH)}}$
is the maximal value of $R$ for the black hole formation.
According to this Table, the result that the black ring horizon
does not form in the $n\le 6$ cases is realistic, because 
$R_{min}^{\textrm{(BR)}}$ is almost constant for all $n\ge 7$ and
$R_{max}^{\textrm{(BR)}}$ linearly decreases with the decrease in $n$
and thus $R_{max}^{\textrm{(BR)}}$ becomes smaller than 
$R_{min}^{\textrm{(BR)}}$ for $n\le 6$ if we extrapolate them.
Because the value of $R_{max}^{\textrm{(BH)}}$ is 
larger than $R_{max}^{\textrm{(BR)}}$ in the cases of $n=7$ and $8$,
the ring horizon is not an apparent horizon because it is 
inside the black hole horizon. The minimum particle number $n_{\rm min}$
for the ring apparent horizon formation is nine, although the parameter
range of $R$ is small in this case. 
Hence in the $D=5$ case, to produce black rings
using particle systems would be extremely difficult.  
Figure~\ref{pic:D5} shows 
the shape of the horizons in the cases of $n=7$ and $n=9$.

\begin{table*}[t]
\caption{\label{data:D7_8}The values of $R_{min}^{\textrm{(BR)}}$, 
$R_{max}^{\textrm{(BR)}}$, and $R_{max}^{\textrm{(BH)}}$ 
for $n=3,..., 7$ in the cases of $D=7$ and $8$. 
In these dimensions, the ring horizon does not form for $n=3$
and the ring apparent horizon forms for $n\ge n_{\rm min}=4$.
The parameter range of $R$ for the black ring formation for $D=8$ is somewhat
larger than the $D=7$ case.}
\begin{ruledtabular}
\begin{tabular}{c|ccccc|ccccc}
 &\multicolumn{5}{c}{$D=7$}&\multicolumn{5}{c}{$D=8$}\\
  $n$ & 3 & 4 & 5 & 6 & 7 & 3 & 4 & 5 & 6 & 7   \\
  \hline 
  $R_{min}^{\textrm{(BR)}}$ & -- & 0.89 & 0.90 & 0.91 & 0.91 
         & -- & 0.91 & 0.92 & 0.92 & 0.92 \\
  $R_{max}^{\textrm{(BR)}}$ & -- & 0.96 & 1.10 & 1.24 & 1.38 
         & -- & 1.03 & 1.19 & 1.35 & 1.51 \\
  $R_{max}^{\textrm{(BH)}}$ & 0.85 & 0.88 & 0.90 & 0.90 & 0.90 
         & 0.87 & 0.90 & 0.91 & 0.91 & 0.91 \\
\end{tabular}
\end{ruledtabular}
\end{table*}

\begin{figure}[t]
\centering
{
\includegraphics[height=0.25\textheight]{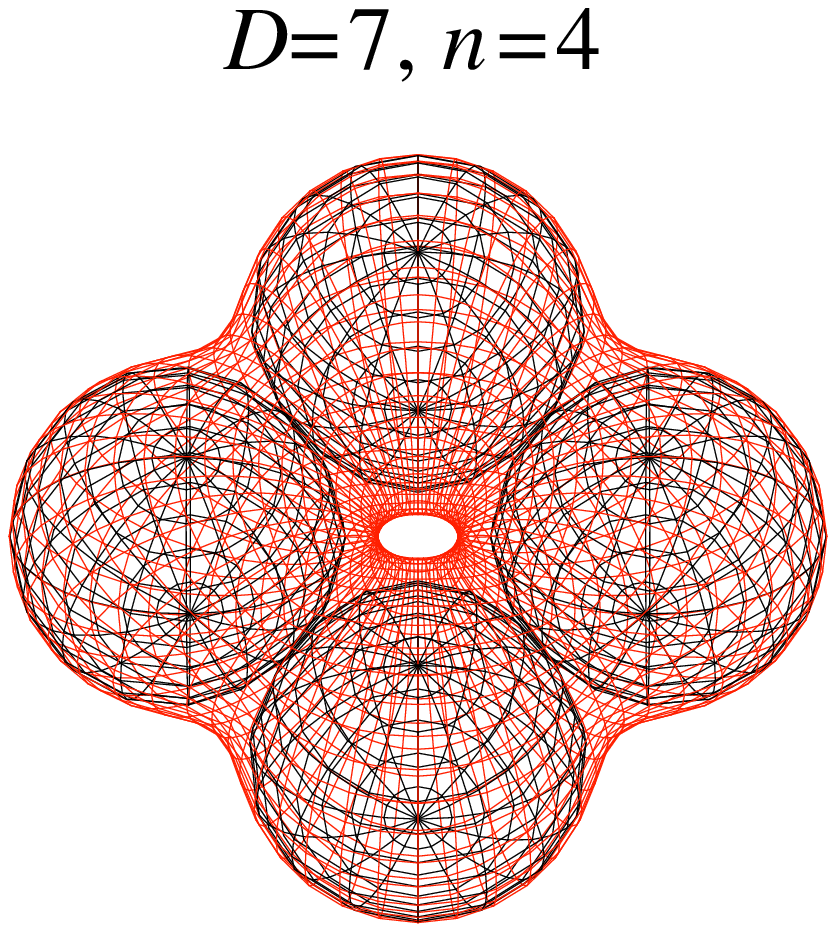}
}
\caption{\label{pic:D7}The black ring horizon in a four-particle system in
seven-dimensions ($R=0.9\tilde{r}_h$).}
\end{figure}

We turn to the $D=6$ case. 
Table~\ref{data:D6} shows the values of $R_{min}^{\textrm{(BR)}}$,
$R_{max}^{\textrm{(BR)}}$, and $R_{max}^{\textrm{(BH)}}$
for $n=3,...,7$. In the $D=6$ case, the ring horizon forms for $n\ge 4$.
Because it is enclosed by the black hole horizon in 
the $n=4$ case,
the ring apparent horizon appears only for $n\ge n_{\rm min}=5$. This value is
far smaller compared to the $D=5$ case. At the same time, the parameter
range of $R$ that the black ring horizon appears is far larger than the $D=5$
case if we compare fixing $n$.
Hence we find that the ring horizon formation becomes easier 
for larger $D$.
The shape of the horizons in the $n=4$ and $n=5$ cases are shown 
in Fig.~\ref{pic:D6}.

Now we show the results of $D=7$ and $8$, which 
are summarized in Table~\ref{data:D7_8}. 
In the $D=7$ case, the ring horizon does not form in a 
three-particle system. For $n\ge n_{\rm min}=4$, the black ring horizon 
that is not enclosed by the black hole horizon appears.
The result for $D=8$ is similar to the $D=7$ case, except
that the parameter range of $R$ of the black ring formation
becomes somewhat larger than the $D=7$ case if we compare
fixing $n$. We find again that the black ring becomes
easier to form for larger $D$. The ring horizon 
in a four-particle system in the $D=7$ case is shown in
Fig.~\ref{pic:D7}.

\begin{table*}[t]
\caption{\label{data:D9_10}The values of $R_{min}^{\textrm{(BR)}}$, 
$R_{max}^{\textrm{(BR)}}$, and $R_{max}^{\textrm{(BH)}}$ 
for $n=3,..., 7$ in the cases of $D=9$ and $10$. 
The ring apparent horizon forms even in the three-particle system
in these dimensions.}
\begin{ruledtabular}
\begin{tabular}{c|ccccc|ccccc}
 &\multicolumn{5}{c}{$D=9$}&\multicolumn{5}{c}{$D=10$}\\
  $n$ & 3 & 4 & 5 & 6 & 7 & 3 & 4 & 5 & 6 & 7   \\
  \hline 
  $R_{min}^{\textrm{(BR)}}$ & 0.90 & 0.93 & 0.93 & 0.94 & 0.94 
         & 0.92 & 0.94 & 0.94 & 0.95 & 0.95 \\
  $R_{max}^{\textrm{(BR)}}$ & 0.92 & 1.08 & 1.25 & 1.43 & 1.60 
         & 0.94 & 1.11 & 1.30 & 1.48 & 1.67 \\
  $R_{max}^{\textrm{(BH)}}$ & 0.89 & 0.92 & 0.92 & 0.92 & 0.93 
         & 0.91 & 0.93 & 0.93 & 0.93 & 0.93 \\
\end{tabular}
\end{ruledtabular}
\end{table*}

\begin{figure}[t]
\centering
{
\includegraphics[height=0.25\textheight]{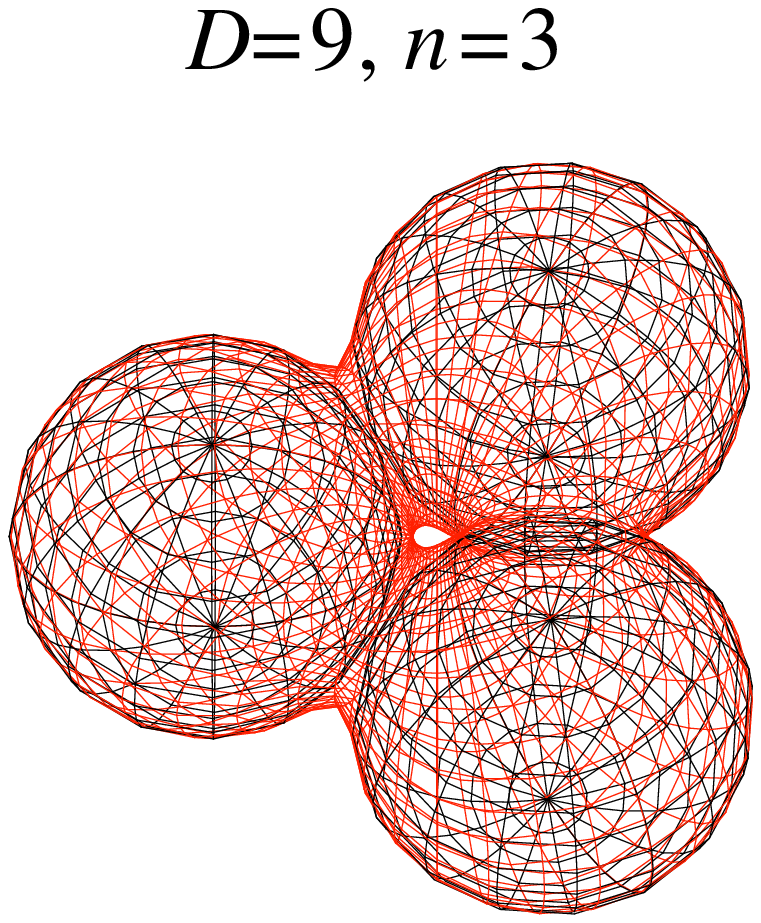}
}
\caption{\label{pic:D9}
The black ring horizon in a three-particle system
in nine-dimensions ($R=0.91\tilde{r}_h$).}
\end{figure}

Finally we look at the $D=9$ and $D=10$ cases.
The results are summarized in Table~\ref{data:D9_10}.
In these cases, the black ring formation is possible 
even in the three-particle system, although the parameter
range of $R$ is quite small. The shape of the black ring horizon
in a three-particle system in nine-dimensions are
shown in Fig.~\ref{pic:D9}. 
We summarized the minimum particle number $n_{\rm min}$ that is required
for the ring apparent horizon formation for each $D$ in Table~\ref{minimum_n}.

\begin{table}[t]
\caption{ \label{minimum_n}The minimum particle number $n_{\rm min}$
that is necessary for the black ring formation for each $D$.}
\begin{ruledtabular}
\begin{tabular}{c|cccccccc}
  $D$ & 4 & 5 & 6 & 7 & 8 & 9 & 10 & 11 \\
  \hline 
  $n$ & -- & 9 & 5 & 4 & 4 & 3 & 3 & 3 \\
\end{tabular}
\end{ruledtabular}
\end{table}

\begin{figure}[t]
\centering
{
\includegraphics[width=0.6\linewidth]{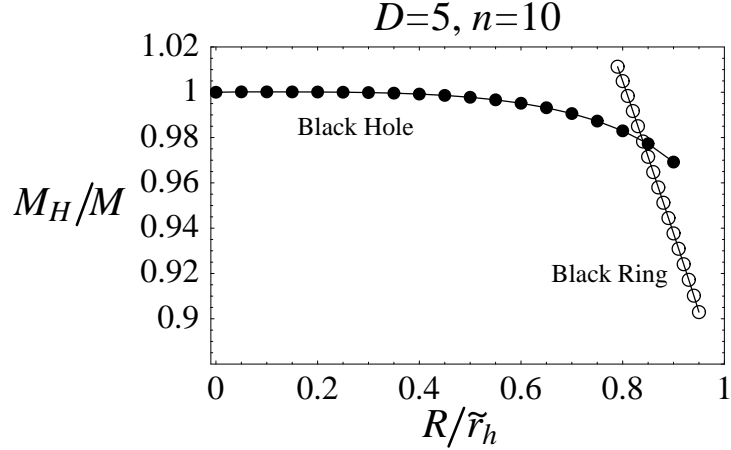}
}
\caption{\label{area:D5n10}
The behavior of Hawking's quasi-local mass $M_{H}^{\rm (BH)}$ ($\bullet$)
and $M_{H}^{\rm (BR)}$ ($\circ$) on the horizons 
as functions of $R$ in the ten-particle
system in five-dimensions. Both $M_{H}^{\rm (BH)}$ and
$M_{H}^{\rm (BR)}$ are the monotonically decreasing functions of $R$.
The value of $M_{H}^{\rm (BR)}$ is greater than the ADM mass at 
$R=R_{min}^{(BR)}$, but it rapidly decreases and it is smaller than
$M_{H}^{\rm (BH)}$ at $R=R_{max}^{\rm (BH)}$. The value of the apparent horizon
mass $M_{AH}$ at $R=R_{max}^{\rm (BR)}$ is about 90\% of the ADM mass.
}
\end{figure}

\begin{figure}[t]
\centering
{
\includegraphics[width=0.6\linewidth]{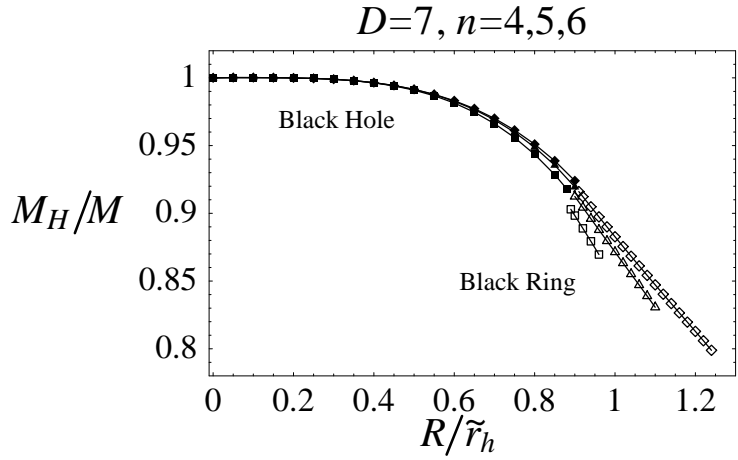}
}
\caption{\label{area:D7}
The behavior of Hawking's quasi-local mass $M_{H}^{\rm (BH)}$ for
$n=4$ ($\blacksquare$), $5$ ($\blacktriangle$), $6$ ($\blacklozenge$)
and $M_{H}^{\rm (BR)}$ for 
$n=4$ ($\square$), $5$ ($\vartriangle$), $6$ ($\lozenge$) 
as functions of $R$ in the $D=7$ case. The values of 
$R_{min}^{\rm (BR)}$ and $R_{max}^{\rm (BH)}$ are similar
and $M_H^{\rm (BR)}$ begins at $R=R_{min}^{\rm (BR)}$ with 
a similar value to $M_H^{\rm (BH)}$.
Similarly to the $D=5$ case, $M_{AH}$ is a monotonically
decreasing function that is discontinuous at $R=R_{max}^{\rm (BH)}$.
}
\end{figure}

Now we calculate Hawking's quasi-local mass~\cite{H68}
on the horizons to understand the amount of trapped energy. 
It is given by
\begin{equation}
M_{H}\equiv \frac{(D-2)\Omega_{D-2}}{16\pi G_D}
\left(\frac{A_{D-2}}{\Omega_{D-2}}\right)^{(D-3)/(D-2)},
\end{equation}
on a surface with zero expansion and becomes an indicator for the energy
trapped by the surface. If the surface is an apparent horizon,
Hawking's quasi-local mass becomes the apparent horizon mass $M_{AH}$,
that gives the lower bound of the mass of the final state
as we see from the area theorem. 
Figure~\ref{area:D5n10} shows the values of Hawking's quasi-local mass on the
black hole horizon $M_H^{\rm (BH)}$ and the black ring horizon 
$M_H^{\rm (BR)}$ of the ten-particle system in the five-dimensional case. 
The value of $M_H^{\rm (BH)}$ is $M$ at $R=0$ because it is the Schwarzschild
spacetime, and it monotonically
decreases as $R$ increases. $M_H^{\rm (BR)}$
has its value in the range $R_{min}^{\rm (BR)}\le R\le R_{max}^{\rm (BR)}$.
Although the value of $M_H^{\rm (BR)}$ is greater than $M$ 
at $R=R_{\min}^{\rm (BR)}$, it rapidly decreases with the increase in $R$
and becomes smaller than $M_{H}^{\rm (BH)}$ at $R=R_{max}^{\rm (BH)}$.
Hence the apparent horizon mass $M_{AH}$ is always smaller than
the ADM mass, and it is a monotonically decreasing function of $R$
that is defined in $0\le R\le R_{max}^{\rm (BR)}$ and is discontinuous
at $R=R_{max}^{\rm (BH)}$. The shapes of $M_H^{\rm (BH)}$ 
of different $n$ values almost coincide. 
The shapes of $M_H^{\rm (BR)}$ of different $n$ values are almost similar
in the range their values exist: the functions $M_H^{\rm (BR)}$
start at almost the same points and
show the similar dependence on $R$, although
their end points $R=R_{max}^{\rm (BR)}$ are different each other. 
The minimal value of $M_{AH}$ (i.e. $M_H^{\rm (BR)}$
at $R=R_{max}^{\rm (BR)}$) decreases as $n$ increases. 
In the limit $n\to \infty$, the value of $R_{max}^{\rm (BR)}$
becomes arbitrarily large while the minimal value of $M_{AH}$ 
asymptotes to zero, 
because this system appraoches to the ring configuration
analyzed in \cite{IN02}. 
Figure~\ref{area:D7} shows the behavior of Hawking's quasi-local mass 
on the horizons in the $D=7$ case. 
For $D\ge 6$, the parameter range of the overlapping region
$R_{min}^{\rm (BR)}\le R\le R_{max}^{\rm (BH)}$ 
that both the black hole horizon
and the black ring horizon exist becomes small
and this tendency is enhanced for larger $D$. 
In the $D\ge 6$ cases, 
the value of $M_H^{\rm (BR)}$ at $R=R_{min}^{\rm (BR)}$
is slightly less than $M_H^{\rm (BH)}$ for all $n$. 
Similarly to the $D=5$ case, the apparent horizon mass $M_{AH}$
is a monotonically decreasing function of $R$ that is discontinuous
at $R=R_{max}^{\rm (BH)}$. 
If we compare the same $n$ cases, the minimal values of $M_{AH}$  
are similar for all $D\ge 7$. 
It is about 85\%, 82\%, 79\%, and 77\% of the ADM mass for $n=4,5,6$, and $7$
respectively. In the $D=6$ cases, their values are somewhat larger than these 
values. In the three-particle systems in $D=9$ and $10$, 
the value of $M_H^{\rm (BR)}$ is about 87\% of the ADM mass. 
In summary, the amount of energy  
that is trapped by the ring horizon is larger
than 75\% of the ADM mass for all $n\le 7$ cases, 
which can be regarded as a fairy large value. 

\subsection{Physical interpretation}

Now we discuss the physical interpretation for the minimum
particle number $n_{\rm min}$ that is required for
the ring apparent horizon formation. One plausible
interpretation is as follows. Suppose
that $n$ point sources with mass $M/n$ are put
at $n$ vertexes of a regular polygon that is inscribed in a
circle with radius $R$. Each point source has a trapped
region with radius $r_h(M/n)$, where we used the expression 
of eq.~\eqref{eq:Schwarzschild_radius}. The black ring can be expected
to form when the trapped regions of neighboring particles overlap each other.
This condition leads to
\begin{equation}
R\sin({\pi}/{n})<r_h(M/n).
\label{plausible1}
\end{equation}
On the other hand, this $n$-particle system can be 
expected to become a black hole when
all the particles are within a sphere with the gravitational
radius of the system mass. Hence for the ring apparent horizon
formation,  
\begin{equation}
r_h(M)<R
\label{plausible2}
\end{equation}
is also necessary. For the existence of $R$ that
satisfies these two conditions \eqref{plausible1} and \eqref{plausible2}, 
$n$ should satisfy
\begin{equation}
\sin({\pi}/{n})<\left({1}/{n}\right)^{1/(D-3)}.
\end{equation}
The minimum number $n\equiv n_1$ that satisfies this
inequality is summarized in Table \ref{minimum_n_volume}.
Although these values are slightly larger than the 
values derived by the numerical calculations, this interpretation
gives the similar values. The reason why $n_1$ depends on $D$
is that the ratio of $r_h(M/n)$ and $r_h(M)$ has a characteristic value
for each $D$. This is in turn because the each dimension number $D$
has the characteristic power of the gravitational potential.
This leads to the dependence of $n_1$ (and hence 
the minimum number $n_{\rm min}$)
on the dimension number $D$. 
We also would like to mention that the boundary surface 
of the union of particles' trapped
regions does not have $S^1\times S^{D-3}$ topology
for $R<r_h(M/n)$. In this situation, the black ring horizon is expected
to disappear. The overlapping region where the black ring horizon
and the black hole horizon both exist is given by $r_h(M/n)\le R\le r_h(M)$.
This parameter range becomes smaller for larger $D$ if we compare fixing $n$.
All these pictures correspond to the qualitative behavior of the
horizons of our numerical results.

\begin{table}[tb]
\caption{ \label{minimum_n_volume}The minimum particle numbers 
$n_1$ and $n_2$
that are necessary for the black ring formation 
for each $D$ estimated
by our plausible interpretation and Ida and Nakao's conjecture~\cite{IN02}
respectively.}
\begin{ruledtabular}
\begin{tabular}{c|cccccccc}
  $D$ & 4 & 5 & 6 & 7 & 8 & 9 & 10 & 11 \\
  \hline 
  $n_1$ & -- & 10 & 6 & 5 & 4 & 4 & 4 & 3 \\
  $n_2$ & -- & 10 & 9 & 8 & 9 & 9 & 9 & 10 \\
\end{tabular}
\end{ruledtabular}
\end{table}

Here we would like to discuss Ida and Nakao's conjecture~\cite{IN02}
which is the generalization of the hoop conjecture~\cite{Th72} 
in four-dimensional spacetimes.  
The hoop conjecture gives the condition for the apparent horizon
formation in the form $C\le 2\pi r_h(M)$ where $C$ is the characteristic
one-dimensional length scale of the system. This conjecture essentially
states that the mass concentration in all directions is necessary
for the apparent horizon formation, and is
well-confirmed in various systems 
(see \cite{YNT02} and the references therein). 
Ida and Nakao investigated various four-dimensional
conformally-flat momentarily-static spaces in five-dimensions
and found that arbitrarily long black holes can form. 
This shows that one-dimensional length does not provide
the condition for the horizon formation in five-dimensions.
They discussed that the $(D-3)$-dimensional volume $V_{D-3}$ characteristic
to the system would provide the condition for the horizon formation
in $D$-dimensional spacetimes in the form  
$V_{D-3}\lesssim \Omega_{D-3}r_h^{D-3}(M)$.  
In our previous paper~\cite{YN03}, we discussed the validity of this
conjecture in the high-energy two-particle system and found that
this conjecture provides a good condition for the black hole production.

Using our results in this paper, 
we discuss whether this conjecture gives the condition for 
the black ring formation in $n$-particle systems.
In our opinion, this conjecture has a possibility to provide
the condition for the existence of horizons with various topology
if we evaluate the $(D-3)$-volume appropriately.
Here, we assume that the condition of 
the black hole formation is given by $(D-3)$-volume $V^{\rm (BH)}_{D-3}$ 
of a surface with $S^{D-3}$ topology, 
and that the condition of the black ring formation is
given by $(D-3)$-volume $V^{\rm (BR)}_{D-3}$ 
of a surface with $S^1\times S^{D-4}$ topology. 
The value of $V^{\rm (BR)}_{D-3}$ 
in the $n$-particle system is estimated as follows. 
Because the characteristic radius
around $S^{D-4}$ is given by $r_h(M/n)$ and
that around $S^{1}$ becomes $R$, the characteristic 
$(D-3)$-volume becomes 
$V^{\rm (BR)}_{D-3}\sim 2\pi R\times \Omega_{D-4}r_h^{D-4}(M/n)$.
The conjecture gives one condition
\begin{equation}
2\pi \Omega_{D-4}r_h^{D-4}(M/n)R\lesssim \Omega_{D-3}r_h^{D-3}(M).
\label{volume}
\end{equation} 
Because the characteristic $(D-3)$-volume of a surface
with $S^{D-3}$ topology is $V^{\rm (BH)}_{D-3}\simeq \Omega_{D-3}R^{D-3}$,
the system would become a black hole for $R\lesssim r_h(M)$
according to the conjecture.
Hence we should assume one more condition
\begin{equation}
r_h(M)\lesssim R,
\label{split}
\end{equation}
for the black ring apparent horizon formation. 
For the existence of $R$ that
satisfies these two conditions \eqref{volume} and \eqref{split},
the particle number should satisfy
\begin{equation}
n\gtrsim \left(\frac{2\pi\Omega_{D-4}}{\Omega_{D-3}}\right)^{(D-3)/(D-4)}.
\label{IN_estimate}
\end{equation}
The minimum number $n_2$ satisfying this condition~\eqref{IN_estimate}
for each $D$ is shown in Table~\ref{minimum_n_volume}.
These values don't coincide with the values 
in Table~\ref{minimum_n} especially in the point that
$n_2$ is not a monotonically decreasing function of $D$ and
cannot explain the fact that $n_{\rm min}$ approaches to
three for large $D$.
This result might be because our evaluation of the $(D-3)$-volume
was not appropriate, or might show that the coefficient $\Omega_{D-3}$ in 
the inequality $V_{D-3}\lesssim \Omega_{D-3}r_h^{D-3}(M)$
should be taken another appropriate value to judge the existence
of the ring horizon. 
However, this conjecture can explain the existence
of the minimum particle number $n_{\rm min}$ 
within the difference by a factor four for $5\le D\le 11$
even in such a rough estimation. 
Because the aim of this conjecture would be to provide the phenomenological
intuition and would not be to give the exact theorem for the horizon formation,
this result can be regarded as successful.
Furthermore, because the surface 
of which we evaluated the $V^{\rm (BR)}_{D-3}$
value does not have $S^1\times S^{D-4}$ topology
for $R\le r_h(M/n)$, the black ring is expected 
to disappear in this configuration. Hence the conjecture
naturally explains the non-existence of the black ring horizon
for small $R$. 
In these meanings, Ida and Nakao's conjecture provides the
condition for the horizon formation even if its topology is not spherical.

\section{Summary and Discussion}

In this paper, we investigated the black ring formation
in particle systems. In Sec. II, we analyzed the
black ring horizon in the high-energy particle collision.
In the high-energy two-particle system, we explicitly proved that
there is no black ring horizon on the union of two 
incoming shock waves. Even in the high-energy $n$-particle system,
we have seen that there is no ring horizon on the union
of $n$ incoming shock waves. These results indicate that
the black ring horizon does not exist at the instant of 
collision, but they don't imply that the black ring 
does not form in high-energy particle systems: there remains the possibility
of the black ring formation after the collision of incoming shocks.
To understand whether the black ring forms in the high-energy 
particle system or not,
we should investigate the spacetime slice after the collision
where the particles interact each other. However, to obtain 
the metric in this phase is difficult.

As a simplified model of $n$-particles that we can calculate, 
we considered the momentarily-static $n$-black-hole system
originally developed by Brill and Lindquist to
obtain some indications about the feature of the black ring formation
in particle systems.
We numerically solved the black ring horizon and the black hole
horizon that surround $n$ particles. 
We located the $n$ particles at $n$ vertexes of a regular polygon
that is inscribed in a circle with the radius $R$.
There is the minimal value $R_{min}^{\rm (BR)}$ 
and the maximal value $R_{max}^{\rm (BR)}$ such that
the black ring horizon exists only in the range 
$R_{min}^{\rm (BR)}\le R\le R_{max}^{\rm (BR)}$.
For the black ring formation, the distance between the neighboring
particles should be small, while the space at the center
between all the particles should be large. These lead to the
existence of $R_{min}^{\rm (BR)}$ and $R_{max}^{\rm (BR)}$.
For each $D$, there is the minimum particle number $n_{\rm min}$ 
that is necessary
for the ring apparent horizon formation. The results are summarized in 
Table~\ref{minimum_n}. This minimum number $n_{\rm min}$ becomes small
and approaches to three for large $D$. If we compare fixing $n$, the parameter
range $R$ of the black ring formation becomes larger as $D$ increases.
These results show that the black ring formation becomes easier
for larger dimension number $D$. 
Because there is a plausible interpretation for our results,
we expect that these features would be valid also 
in the high-energy particle system.

We calculated Hawking's quasi-local mass on the horizons to understand
the energy trapped by the horizon. Hawking's quasi-local mass 
on the apparent horizon $M_{AH}$ (the apparent horizon mass)
is a monotonically decreasing function of $R$ that is discontinuous
at $R=R_{max}^{\rm (BH)}$ where the black hole horizon disappears. 
The amount of the energy trapped inside the ring horizon is always greater than
75\% for the $n \le 7$ cases, which is rather a large value.
We don't consider that so much energy can be trapped in the case of the 
black ring horizon in the high-energy particle system.
In our previous paper~\cite{YN03}, we calculated the apparent horizon
mass of the black hole horizon that is formed 
in the high-energy two-particle grazing collision.
The energy trapped by the horizon is quite small if $D$ is large and
the impact parameter is close to its maximal value, and we discussed
that many black holes with small mass would be produced in high-energy
particle systems, although further investigations of the gravitational
radiation after the collision would be necessary to obtain
the rigorous answer. Hence we consider that the trapped energy
of the black ring might be small also in the high-energy $n$-particle system.
However we expect that the energy trapped by the black ring
would be always smaller than that trapped by
the black hole if we compare fixing $n$ and $D$ also in the
high-energy $n$-particle system.

From these results, we expect that the production of the black rings
using systems with many particles would be possible, 
although it is difficult
for the $D=5$ case. $n=4$ would be sufficient for
this purpose if $D$ is equal to or larger than seven. We should mention that
the motion of each particle is important to observe
the signals from the black rings in colliders. 
For example, if each particle has a velocity such that the
ring radius decreases, the system would rapidly collapses to 
a black hole and we would not be able to detect the signals 
of the black ring even if the ring horizon momentarily forms.  
To produce the black ring with a long lifetime, each particle
should have a velocity in the $\phi$ direction in the coordinate
system used in Sec. III. If we take it into account, the ordinary
colliders are not appropriate for the black ring formation. A linear
collider with four arms, for example,  would be able to produce the black rings
with a small probability.

Our remaining problem is as follows. 
Although we expect that the basic feature of the results in
the Brill-Lindquist initial data would not be changed 
in the high-energy particle system, there is a possibility that  
the momentums of particles have a large effect
on the existence of the horizon 
as we have shown in our previous paper in
two-black-hole systems in four-dimensions~\cite{YNT02}.
To clarify this point, we should investigate the case where
each particle has a momentum in the $\phi$ direction. 
For this purpose, Bowen and York's method~\cite{BY80} 
is useful. Using this method, one can calculate the conformally-flat
initial data of multi-black holes with momentums. The extrinsic curvature
is given analytically, and one should solve the conformal factor numerically.
Although this method was proposed in the four-dimensional case,
the extension to the higher-dimensional cases has been done by
Shibata~\cite{Shiba02}.  Using this method, we are able to
calculate the initial data of moving particles and understand
the effect of motion on the ring horizon formation.
Next, we would like to investigate the remaining possibility
to produce the black rings in two-particle systems. 
Although it seems that the black ring would not form in the 
two-point-particle system from the results in this paper, 
it might be possible if we consider the deformed particles.
For example, if we regard the incoming particle as the fundamental string,
the classical model for the gravitational field would be provided
by a segmental source. Hence in the high-energy collision of
two-strings, there is a possibility of the black ring formation. 
Another example uses the instability of the higher-dimensional 
Kerr black holes.  Recently
Emparan and Myers have shown that the ultra-spinning Kerr black holes
in higher-dimensions are unstable~\cite{EM03}. Hence, if these ultra-spinning
Kerr black holes can be produced in the high-energy two-particle collisions,
they would deform due to this instability.
Subsequently two deformed black holes
might collide each other and form the black ring.
To confirm this scenario, we should
investigate whether the production of the ultra-spinning
black holes is possible in the high-energy two-particle system. 
This requires the investigation of the balding phase 
(i.e. the evolution after the particle collision) 
and the determination of the amount of energy and angular momentum
that is radiated away.

\acknowledgments

We would like to thank Masaru Shibata, Ken-ichi Nakao and Daisuke Ida for
helpful discussions. We are also in thank of Tetsuya Shiromizu's 
helpful comments on the black ring formation in the system of
two deformed particles.
The work of H. Y. is supported in part by a grant-in-aid from 
Nagoya University 21st Century COE Program (ORIUM).


\end{document}